# Potential and Challenge of Ankylography

Jianwei Miao[1], Chien-Chun Chen[1], Yu Mao[2], Leigh S. Martin[1,3] and Henry C. Kapteyn[3]
[1]Department of Physics and Astronomy, and California NanoSystems Institute, University of California, Los Angeles, CA 90095, USA. [2]Institute for Mathematics and Its Applications, University of Minnesota, Minneapolis, MN 55455, USA. [3]Department of Physics and JILA, University of Colorado and NIST, Boulder, CO 80309, USA.
Email: miao@physics.ucla.edu.

The concept of ankylography, which under certain circumstances enables 3D structure determination from a single view[1], had ignited a lively debate even before its publication[2,3]. Since then, a number of readers requested the ankylographic reconstruction codes from us. To facilitate a better understanding of ankylography, we posted the source codes of the ankylographic reconstruction on a public website and encouraged interested readers to download the codes and test the method[4]. Those who have tested our codes confirm that the principle of ankylography works. Furthermore, our mathematical analysis and numerical simulations suggest that, for a continuous object with array size of 14x14x14 voxels, its 3D structure can usually be reconstructed from the diffraction intensities sampled on a spherical shell of 1 voxel thick[4]. In some cases where the object does not have very dense structure, ankylography can be applied to reconstruct its 3D image with array size of 25×25×25 voxels[4]. What remains to be elucidated is how to extend ankylography to the reconstruction of larger objects, and what further theoretical, experimental and algorithm developments will be necessary to make ankylography a practical and useful imaging tool. Here we present our up-to-date understanding of the potential and challenge of ankylography. Further, we clarify some misconceptions on ankylography, and respond to technical comments raised by Wei[5] and Wang *et al.*[6] Finally, it is worthwhile to point out that the potential for recovering 3D information from the Fourier coefficients within a spherical shell may also find application in other fields.

Wei provided a mathematical proof to argue that single-shot diffractive imaging of truly 3D structures suffers from a dimensional deficiency and does not scale[5]. A critical assumption made in his proof is the spherical shell to be infinitesimal thin. However, ankylography is based on the grid points within a spherical shell of 1 voxel thick, as stated in the Supplementary Information to Raines *et al.*[1]. Experimentally, the finite thickness of the spherical shell can be achieved by controlling the energy bandwidth and the divergence/convergence angle of the incident beam, while the infinitesimally thin shell requires the bandwidth ($\Delta E$) and the divergence/convergence ($\Delta\theta$) to be zero which cannot be realized in real experiments. Mathematically, it can be shown via matrix rank analysis[7,8] that the sampling matrix of the grid points on an infinitesimally thin shell does not have full rank no matter how small the tolerance is specified. The mathematical arguments in Wei and Wang *et al.* draw a similar conclusion for the grid points on an infinitesimally thin shell[5,6]. However, for the grid points within a 1 voxel thick shell, our matrix rank analysis indicates that, if the tolerance is small enough, the sampling matrix has full rank[7,8]. When the 3D array of an object gets larger, the tolerance becomes smaller in order to maintain full rank of the sampling matrix. When the tolerance approaches to zero, the sampling matrix becomes very difficult or impossible to be inverted (an ill-conditioned problem) and the inversion also becomes extremely sensitive to noise. This is why ankylography currently cannot be used to reconstruct larger objects without more sophisticated reconstruction strategies, additional information and constraints. Although Wei has done a thorough analysis of the dimensional deficiency and uniqueness of the grid points on an infinitesimally thin shell, he did not perform any analysis for the grid points within a 1 voxel thick shell[5]. This may explain why Wei's conclusion - "SSDI in general suffers from a dimensional deficiency that limits the applicability of ankylography to objects that are small-sized in at least one dimension or that are essentially 2D otherwise."- is inconsistent with our results. Both our numerical simulations and experiments suggest that ankylography work well for 3D (not just 1D or 2D) objects[1,4,8], although the array size is currently limited. Furthermore, Eqs. 1 and 2 in Wang *et al.*[6] suggest that the Fourier transform of a function, $g(\vec{r})$, sampled on an infinitesimally thin shell can in principle be all zeros. But, this is experimentally impossible for any real, finite objects due to the finite $\Delta E$ and $\Delta \theta$ of the incident beam.

In Raines *et al.*[1], two numerical simulations were presented. The first was for a sodium silicate glass particle with array size of 14x14x14 voxels, which was directly reconstructed from the oversampled diffraction intensities within a spherical shell of 1 voxel thick. The other is a poliovirus with array size of 32x32x20 voxels in which both the 3D capsid and interior structure of the virus are reconstructed. In the latter case, amplitude extension was developed to assist the ankylographic reconstruction. Wang *et al.* simulated an object of 64x64x64 voxels[6], in which the size along in the Z-axis (*i.e.* the critical direction) is more than three times larger than what we attempted. Thus it is not surprising that they failed in the simulation. Incidentally, we have also sent them our ankylographic reconstruction code and they have confirmed it works for a small array. Wang *et al.* also compared ankylography with "divergent tomography"[6]. Although the two methods may have some similarity, they are fundamentally different. Ankylography is based on *coherent diffraction*, where each intensity point is a function of *all* the electron density points of the 3D object. But tomography is an *incoherent imaging* method where each detector element is a line integral of the absorption coefficients inside an object. The difference between ankylography and divergent tomography is analogous to the difference between crystallography and tomography. Furthermore, oversampling is only applicable to reciprocal space (*e.g.* ankylography), but not to real space (*e.g.* divergent tomography). Finally, Wei questioned how Poisson noise was added to the diffraction patterns in Raines *et al.*[1]. With a given incident flux and a known 3D structure, the calculation of Poisson noise in an oversampled diffraction pattern is well-known in the CDI community.

Compared to CDI[9], ankylography, which recovers 3D structure information from oversampled diffraction intensities on a spherical shell, is an ill-posed problem and inherently not a general method. However, an ill-posed problem does not mean it cannot be practically dealt with. For example, super-resolution image reconstruction is a field related to ankylography, and is also well-known to be ill-conditioned. Nevertheless the super-resolution problem is a popular topic in applied mathematics and several practical techniques has been developed to solve it[10,11]. Furthermore, 3D structure may also be determined from 1D atomic pair distribution function[12] or powder diffraction patterns[13] through the use of constraints. Both intend to reconstruct a 3D structure from a 1D diffraction intensity distribution (*i.e.* an ill-posed problem), which is in principle more difficult than ankylography.

In order to extend ankylography to larger objects, we have been investigating two potential approaches[8]. First, our recent simulation studies suggest that increasing the shell thickness can improve the ankylographic reconstruction of larger objects. Experimentally, this can be realized by using an incident beam with an energy bandwidth, coupled with an energy-resolved detector. Such energy resolved pixel arrays are under development based on superconducting transition edge sensors[14]. Second, by acquiring several spherical diffraction patterns at different sample orientations, ankylography can in principle be extended to larger objects. Fig. 1 shows ankylographic reconstructions of a portion (31x31x31 voxels) of a *drosophila* neuron by using 2, 3 and 4 spherical diffraction patterns. Our simulation results suggest, compared to conventional 3D imaging methods such as tomography, ankylography require much fewer number projections due to the use of spherical diffraction patterns. Since the ankylographic reconstruction utilizes data points sampling on a regular grid, projecting intensity points from the planner detector to a regular grid requires interpolation and introduces artefacts, which remains an important issue and needs further investigation.

As ankylography is such a new idea, it is certainly impossible for us to envision and address all the problems in the first paper. In order to fully understand the potential and limits of ankylography, follow-up studies are needed in theory, experiment and algorithm development. For example, two recent experiments have been reported to further confirm ankylography[8,15]. Finally, it is also worthwhile to point out that, after Raines *et al.*[1], two more related papers have been published in Nature. The first is super-resolution biomolecular crystallography[16], which under some conditions can determine the high-resolution 3D structure of macromolecules from low-resolution data. The other is discrete tomography[17], which enables to achieve the 3D atomic reconstruction of a small crystalline nanoparticle by only using two projections, combined with prior knowledge of the particle's lattice structure. These three papers share common features: i) mathematically ill-conditioned problems; ii) inherently not general methods; and iii) retrieving 3D structural information from a portion of Fourier magnitudes or coefficients. While many issues remain to be resolved, they represent a new and potentially important direction in 3D structural determination.

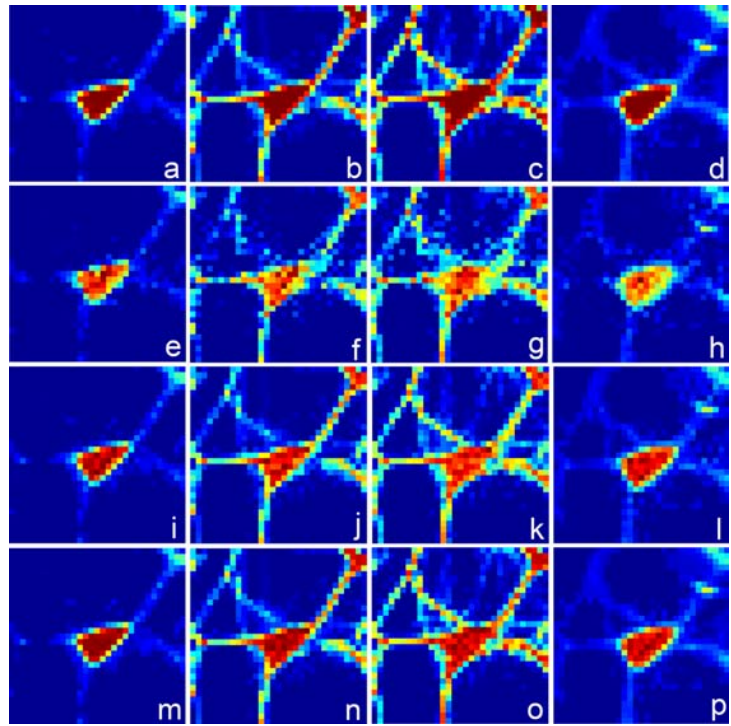

**Figure 1.** Numerical simulations on ankylographic reconstructions of a portion (31x31x31 voxels) of a *drosophila* neuron using 2, 3 and 4 spherical diffraction patterns. The oversampling degree ($O_d$) of these reconstructions is 2.02 and the number of iterations is $5\times10^4$. **a-d**, Slices 10, 15, 20 and 25 of the portion of the 3D neuron structure. **e-h**, The corresponding slices reconstructed from 2 spherical diffraction patterns. The tilt angles for the 2 spherical patterns are -45° and +45°, and the thickness of each spherical shell is 1 voxel. **i-l**, The corresponding slices reconstructed from 3 spherical diffraction patterns. The tilt angles for the 3 spherical patterns are -60°, 0° and +60°. **m-p**, The corresponding slices reconstructed from 4 spherical diffraction patterns. The tilt angles for the 4 spherical patterns are -67.5°, -22.5°, +22.5° and +67.5°. The size of the portion of the neuron is about 35 μm.